\documentclass[sigconf,screen]{acmart}
\AtBeginDocument{%
  }

\acmConference[]{}{}{}
\renewcommand\footnotetextcopyrightpermission[1]{} 

\begin{document}

\title{Takeaways from Applying LLM Capabilities to Multiple Conversational Avatars in a VR Pilot Study}

\author{Mykola Maslych}
\affiliation{%
  \institution{University of Central Florida}
  \city{Orlando}
  \state{Florida}
  \country{USA}}
\email{maslychm@gmail.com}

\author{Christian Pumarada}
\affiliation{%
  \institution{University of Central Florida}
  \city{Orlando}
  \state{Florida}
  \country{USA}}
\email{cpuma1824@gmail.com}

\author{Amirpouya Ghasemaghaei}
\affiliation{%
  \institution{University of Central Florida}
  \city{Orlando}
  \state{Florida}
  \country{USA}}
\email{aghaei.ap@ucf.edu}

\author{Joseph J. LaViola Jr.}
\affiliation{%
  \institution{University of Central Florida}
  \city{Orlando}
  \state{Florida}
  \country{USA}}
\email{jlaviola@ucf.edu}

\renewcommand{\shortauthors}{Maslych et al.}

\begin{abstract}
We present a virtual reality (VR) environment featuring conversational avatars powered by a locally-deployed LLM, integrated with automatic speech recognition (ASR), text-to-speech (TTS), and lip-syncing. Through a pilot study, we explored the effects of three types of avatar status indicators during response generation. Our findings reveal design considerations for improving responsiveness and realism in LLM-driven conversational systems. We also detail two system architectures: one using an LLM-based state machine to control avatar behavior and another integrating retrieval-augmented generation (RAG) for context-grounded responses. Together, these contributions offer practical insights to guide future work in developing task-oriented conversational AI in VR environments.
\end{abstract}



\begin{teaserfigure}
  \includegraphics[width=\textwidth]{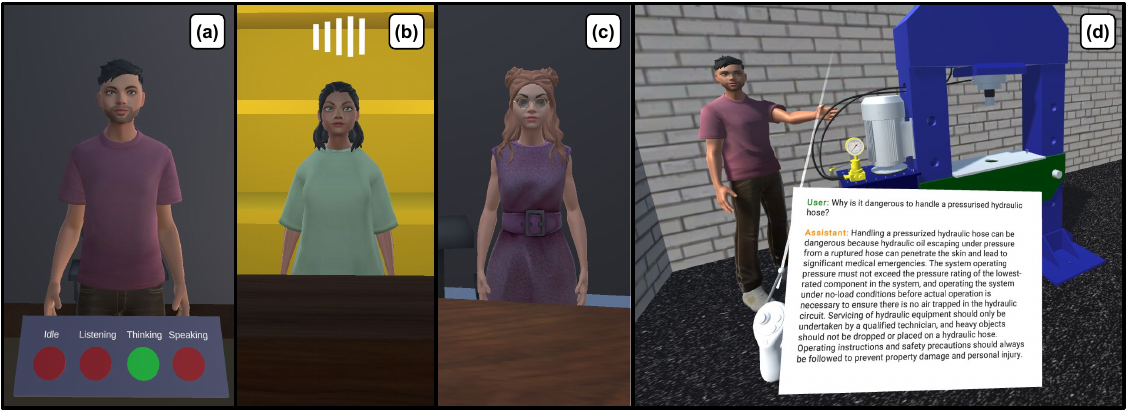}
  \caption{Avatars used in the study with example feedback types: \textbf{(a)}~Friend $+$ \textit{state lights} (one active state among Idle, Listening, Thinking, and Speaking), \textbf{(b)}~Clerk $+$ \textit{loading bar} (appears above avatar only during Thinking state), \textbf{(c)}~Manager $+$ \textit{no feedback} (no processing indication); \textbf{(d)}~On-hand UI with user's query and system response in the industrial training demo application.}
  \Description{Enjoying the baseball game from the third-base
  seats. Ichiro Suzuki preparing to bat.}
  \Description{Figure shows screenshots of the three virtual avatars used in the study, the system state feedback indicators, and a screenshot of a virtual hydraulic press used for the industrial training demonstration. State lights feedback indicator has four states, three of which are red and one is green, indicating the active state; Loading bar indicator appears as a two-dimensional loading bar above the avatar during system response generation; No feedback shows no indication that system is generating a response.}
  \label{fig:teaser}
\end{teaserfigure}


\maketitle

\section{Introduction}

Increased focus on Large Language Models (LLMs) has led to significant improvements in the quality of generated text, facilitating development of task-specific LLMs. Realism of Non-Playable-Characters (NPCs) in consumer applications has benefited from these advancements~\cite{simularca2023}, and in academia, LLM-powered intelligent virtual agents (IVAs) are being applied to learning~\cite{divekar_foreign_2022, wang_virtuwander_2024}, health support~\cite{saeed_developing_2024, wang_designing_2024}, development process~\cite{qin_charactermeet_2024}, and companionship~\cite{zhu_free_form_2023, wan_building_2024}, among other uses. In this preliminary work, we explore how participants behave while vocally conversing with virtual avatars to inform the development of future systems.

\begin{figure*}[!ht]
    \centering
    \includegraphics[width=0.99\linewidth]{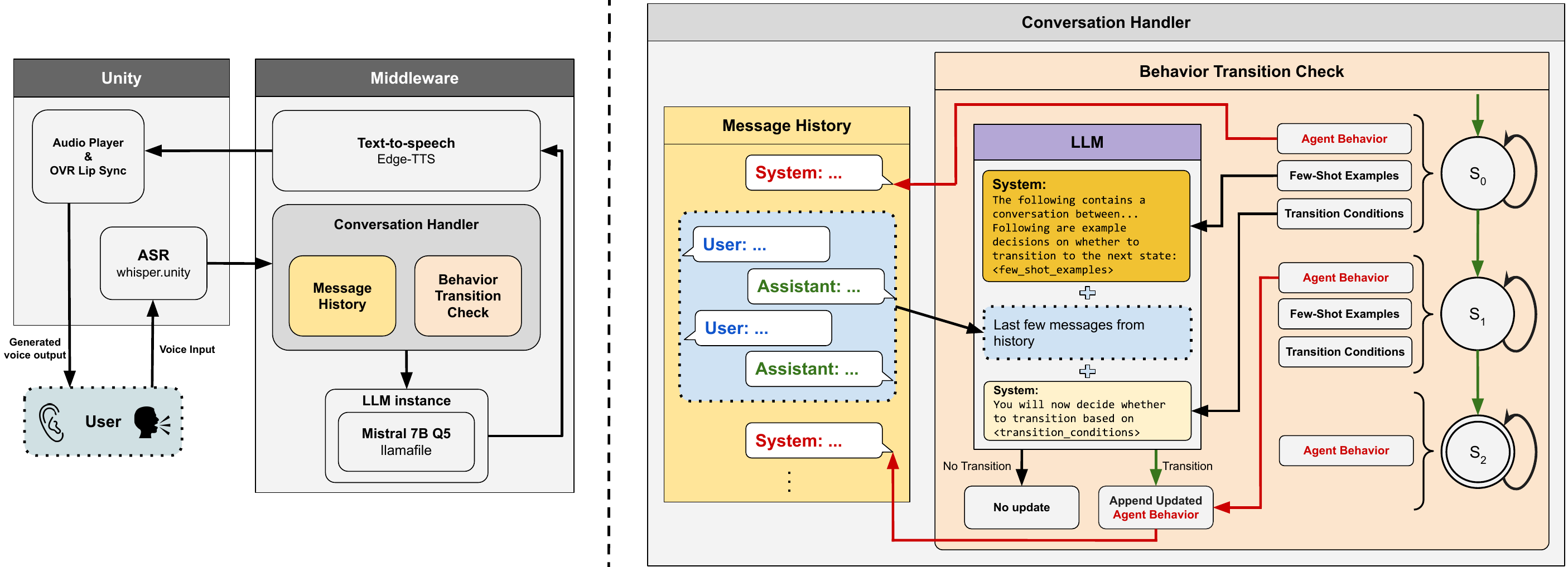}
    \caption{Pipeline for generating responses to user's queries. \textbf{Left -- architecture overview}: ASR transcribes user's voice, passing it to Conversation Handler, which uses an LLM to generate a text response that gets voiced by Edge-TTS. \textbf{Right -- Conversation Handler}: state management system for controlling agent's behavior. Each state contains agent behavior that gets appended as a system message upon a transition to that state; states with outgoing transitions also contain transition conditions and few-shot examples of transition decisions. 
    Transitions are decided by an LLM, which is instructed to return "transition" / "no transition" responses through system prompts with the last few messages from user-avatar history inserted in-between.}
    \label{fig:study_architecture}
    \Description{Figure shows two flowchart diagrams that explain the flow of data in the system to generate a response. Diagram on the left shows that the output of speech recognition is fed to the Mistrag 7B LLM, which generates response text. Speech generation is then applied on the text, and resulting audio is played inside the Quest3 HMD that participant wears, completing the interaction loop. Diagram on the right shows that agent behavior is controlled by appending an extra system message into the LLM history. Decision on whether to append a system message is made by another LLM instance that scans the last few messages between the user and assistant and uses prompts with conditions to determine if a transition should occur.}
\end{figure*}

We developed a system for voice conversational loop powered by a locally-deployed LLM, automatic speech recognition (ASR), and text-to-speech (TTS) through an API. This pipeline was tested in a pilot study where users completed a quest-like scenario by conversing with avatars, whose behavior was controlled by an LLM-based state machine. Further, we created a retrieval-augmented generation (RAG) application, which answers users queries about a digital twin of an industrial machine, generating responses grounded in context extracted from an operation manual. Observing user behavior and collecting system response timings, head gaze directions, and survey responses, provided us with insights into areas of improvement and design of future conversational systems. 

Section~\ref{sec:implementation} covers the architecture and implementation of the multi-agent conversational system, which we used in the pilot study described in \autoref{sec:study}. In \autoref{sec:demo}, we detail the RAG system for a training application, and takeaways with recommendations from working with LLM-based conversational AI are summarized in \autoref{sec:lessons}.

\section{System implementation}
\label{sec:implementation}

\autoref{fig:study_architecture}-Left shows the conversational system architecture. When the system receives an audio input, Whisper~\cite{radford2022robustspeechrecognitionlargescale} Unity package\footnote{\footnotesize\href{https://github.com/Macoron/whisper.unity/}{github.com/Macoron/whisper.unity/}} transcribes it. This transcription is then sent to a middleware server hosted with FastAPI\footnote{\href{https://fastapi.tiangolo.com/}{fastapi.tiangolo.com/}}, which manages message histories of the avatars. The updated message history is passed to the Mistral 7b LLM~\cite{jiang_mistral_2023}, locally hosted with llamafile\footnote{\href{https://github.com/Mozilla-Ocho/llamafile/}{github.com/Mozilla-Ocho/llamafile/}}, which generates a text response from the avatars perspective. This text is then passed to Edge-TTS \footnote{\href{https://github.com/rany2/edge-tts}{github.com/rany2/edge-tts}} system, which generates a voice and saves it as an MP3 file. The path of the audio file is returned to Unity, which downloads it and plays it through a directional Audio Source. OVR Lip Syncing package\footnote{\href{https://developer.oculus.com/documentation/unity/audio-ovrlipsync-unity/}{developer.oculus.com/documentation/unity/audio-ovrlipsync-unity/}} controls the blendshapes on the corresponding avatar's face as the audio was played. The avatars were designed and imported from ReadyPlayerMe package\footnote{\href{https://docs.readyplayer.me/ready-player-me}{docs.readyplayer.me/ready-player-me}}, and the environment was designed to fit the avatars' roles. System response time (SRT), measured between the time when participant finished speaking and the avatar started responding, averaged at 3.2 seconds. 

\subsection{Task Transitions}
\label{sec:transitions}

\begin{table}
\caption{Tasks that appeared on the handheld panel interface. Tasks were crossed-out after they were completed.}
\label{tab:tasks}
\begin{tabular}{ll}
\hline
\multicolumn{1}{c}{\textbf{Task}}             & \multicolumn{1}{c}{\textbf{Appears After}} \\ \hline
(1) Talk to Friend                            & System                                     \\
(2) Buy walnuts from the store                & Friend                                     \\
(3) Bring walnuts to Friend                   & Clerk                                      \\
(4) Ask about milk delivery date at the store & Friend                                     \\
(5) Ask Manager about next shipment date      & Clerk                                      \\
(6) Tell Friend the milk delivery date        & Manager                                    \\ \hline
\end{tabular}
\end{table}

To determine whether the latest task was completed (see tasks in \autoref{tab:tasks}) and if a new one had to be issued, we implemented a state machine through an LLM (see ~\autoref{fig:study_architecture}-Right). Before the system generated avatar's responses, the last few messages between the participant and a current avatar were appended to pre-written system prompts that instructed the LLM to determine whether an event has occured in the conversation (e.g. whether the participant has completed a purchase of walnuts). The LLM outputted a decision in text ("transition" / "no transition"), and in the case of transition, a new system prompt with updated behavior of the avatar was appended to its message history, along with displaying the next task to the participant. 

\begin{figure*}
    \centering
    \includegraphics[width=.99\linewidth]{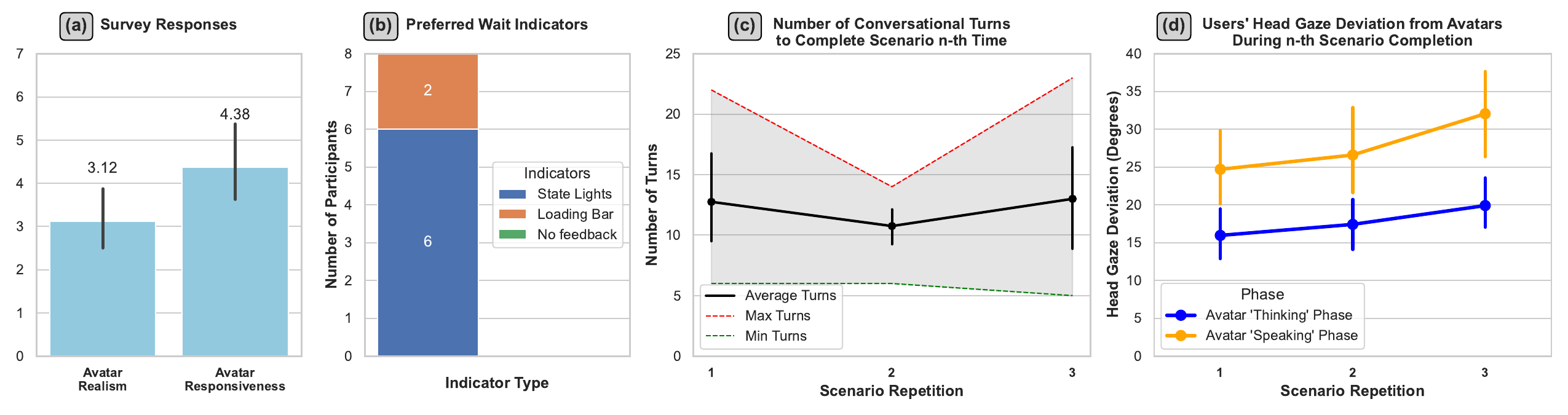}
    \caption{Pilot study results: \textbf{(a)} survey responses about avatar realism and responsiveness; \textbf{(b)} preferred wait feedback types; \textbf{(c)} number of conversational turns required to complete the in-VR scenario n-th time; \textbf{(d)} participant's head gaze deviation angle (from directly looking at the avatar's face) during n-th scenario completion.}
    \label{fig:survey_data}
    \Description{Figure shows four charts. Bar chart A shows that avatar realism ratings averaged at 3.12 out of 7, and avatar responsiveness ratings averaged at 4.38 out of 7. Stacked chart B shows that 6 out of 8 participants selected state lights, and 2 out of 8 participants selected the loading bar as their preferred feedback indicator type. Line chart C shows that number of conversational turns to complete the in-VR scenario varied between the first, second, and the third scenario repetition. Line chart D shows that with each of three scenario repetitions, user's head gaze deviated more and more from directly looking at the avatars' faces.}
\end{figure*}

\section{Pilot Study}
\label{sec:study}

The participants were instructed to navigate a virtual environment in a Meta Quest 3 HMD, completing a scenario with a series of tasks (see~\autoref{tab:tasks}) by speaking with three avatars (Friend, Clerk, Manager) at three different locations (Friend's room, store counter, Manager's office). Each avatar was surrounded by an invisible trigger volume (collider), and when participants entered this volume, the avatar turned its head toward the user. While inside the collider, participants activated voice input with "A" button press on a controller, then pressed it again after they finished speaking. After avatars responded in voice, if a previous task was completed (see~\autoref{sec:transitions}), it would appear as strike-through, and a new task was appended to a text UI attached to the participants' left hand. The first task appeared at the application start. When participants navigated to the Friend's room, the Friend asked them to purchase walnuts from a store. The participants then navigated to the store and talked to the Clerk, completing the purchase task through conversation. After participants brought the walnuts back to the Friend, the new task was to return to the store and ask the Clerk about next milk delivery date. The Clerk told the participants to ask the Manager about the date, and upon completing this, participants returned to the Friend. After informing the Friend about the delivery date, participants took off the HMD and filled out a survey about their experience.

\subsection{Conditions}
\label{sec:conditions}

The participants repeated the scenario three times with three different feedback types: \textit{state lights}, \textit{loading bar}, \textit{no feedback}. The order was counterbalanced using the Balanced Latin Square. 
The \textit{state lights} (\autoref{fig:teaser}-a) highlighted the current interaction stage (Idle = active by default, Listening = audio is being recorded, Thinking = processing, Speaking = avatar is responding).
The \textit{loading bar} (\autoref{fig:teaser}-b) appeared above the avatar's head from the moment the participant pressed the controller button to stop talking, and until the avatar started responding in voice. 
The \textit{no feedback} condition (\autoref{fig:teaser}-c) did not show the current state of the avatar in any way.

\subsection{Results}

Eight participants (6 male, 2 female), aged 18 to 24 participated in our pilot study. Participants rated avatar realism and responsiveness, as well as selected their preferred wait feedback type. Additionally, we recorded the number of conversational turns required to complete the scenario, and collected participants' HMD gaze direction (gaze deviation angle from directly looking at the avatar's face) during avatar's response generation (Thinking) and annunciation (Speaking) phases.

\subsubsection{Survey Responses}

We aggregated the survey data averages into a single plot (\autoref{fig:survey_data}-a) since we found no differences between the three wait feedback conditions. Avatar realism scores were quite low at 3.12 out of 7, which can be explained by the lack of body animations besides lip syncing and avatars turning their head towards the participants. Future studies should include idle and responsive animations, as well as facial expressions to improve realism. Avatar responsiveness was rated more positively than realism (at 4.38 out of 7), still, in future work we will try mitigating the delay caused by SRT (3.2 seconds) through voice and gesture fill-ins, as prior work indicated that such fill-ins can reduce the perceived response latency in related contexts~\cite{kum_can_2022}. While realism and responsiveness were not affected by the wait feedback type, most participants preferred \textit{state lights} (6 out of 8) and the \textit{loading bar} (2 out of 8). Some participants commented that presence of any kind of system processing indication gave them the confidence that the avatar heard them, as compared to no indication at all. This is supported by the fact that no participants selected \textit{no feedback} condition as their preferred one.

\subsubsection{Objective Metrics}

\begin{figure*}[!ht]
    \centering
    \includegraphics[width=.8\linewidth]{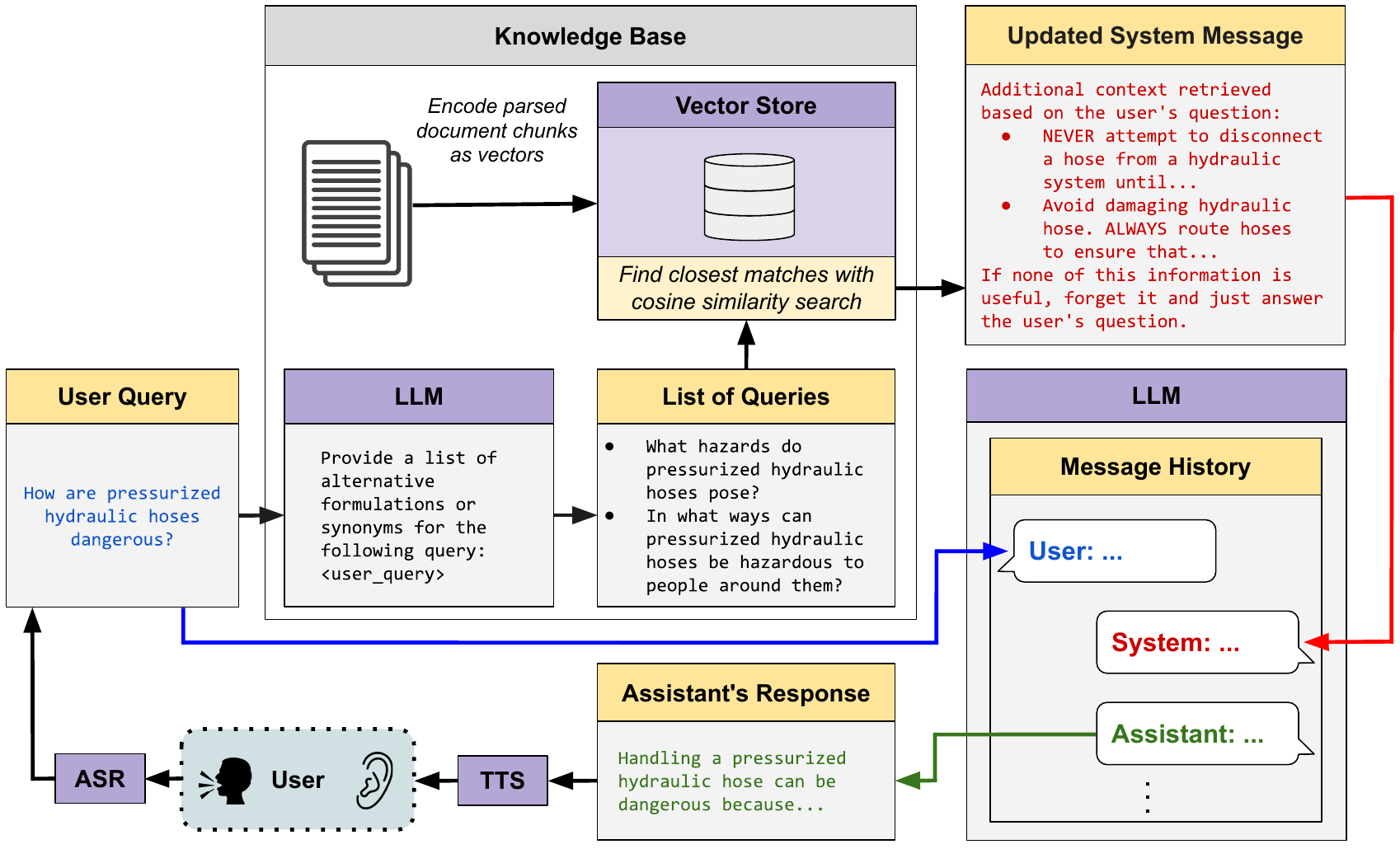}
    \caption{Pipeline for the RAG-enhanced system architecture for answering user's queries about a specific application and machine. After user's speech is transcribed with ASR, alternative formulations of their query are used to retrieve closest matches of text chunks from a machine's manual. This additional context is provided to the LLM as an appended system message.}
    \label{fig:rag_architecture}
    \Description{Figure shows the retrieval augmented generation system pipeline. It adds a knowledge base component, which retrieves sentences from a hydraulic press manual, similar to user's input query. These text chunks are then added to the message history, providing the LLM with context relevant to the user's query.}
\end{figure*}

The average number of conversational turns (see plot in \autoref{fig:survey_data}-c) required to complete the in-VR portion for the first time was $\approx$13, and lowered to $\approx$11 during the second play-through (since participants have learned what to say to the avatars in order to progress in the scenario). However, during the third run, the average increased to $\approx$14 turns, because some participants experimented with the system, testing its limits by saying things unrelated to task completion. Plotting participants' head gaze deviation revealed that participants looked at avatars less and less over the course of scenario repetitions (see \autoref{fig:survey_data}-d). This indicates that in user studies with conversational AI, repeating the same scenario multiple times under varied conditions leads to learning effect (memorization) and lower engagement for some participants, while in others, it leads to undesired experimentation instead of focusing on the completion. Such behaviors add noise to the data and can make detecting differences between conditions more difficult. For a successful user study involving conversational AI with multiple conditions, distinct yet comparable scenarios must be present and counterbalancing carefully applied.

\subsection{RAG Application for Industrial Training}
\label{sec:demo}

We adapted our conversational system to build a demo for an industrial VR training application, where the user could ask questions about a static digital twin of a hydraulic press machine. Unlike system responses for entertainment purposes, responses for safety training must be more precise, so we set the LLM generation temperature to zero~\cite{renze2024effect} and added a RAG component~\cite{wu2024retrievalaugmentedgenerationnaturallanguage}. \autoref{fig:rag_architecture} shows the architecture of the RAG-enhanced system. At application start, a PDF manual for a hydraulic press is parsed into text chunks and encoded as embeddings using sentence transformer~\cite{reimers_2019_sentence_bert}. User queries are reformulated by an LLM, embedded using sentence transformer, and matched to relevant text chunks through cosine similarity search. Before generating a response, a system message with these text chunks is appended to history. In addition to audible output, the text of the latest query and answer was shown on the UI handheld by the user (see \autoref{fig:teaser}-d). 

We demoed this interface during informal showcases, gathering feedback for system improvements and new features. Users appreciated the ability to inspect the 3D machine representation but suggested additional interactivity, such as touch or pointing functionality for targeted queries about machine parts, and a stored per-component message history for revisiting prior queries. Incorporating this feedback, we plan to combine the pipelines in~\autoref{fig:study_architecture} and~\autoref{fig:rag_architecture}, applying them to training~\cite{petersen_pedagogical_2021} and museum exploration studies~\cite{bayat_exploring_2024, wang_virtuwander_2024}.

\section{Discussion}

This section reflects on lessons learned during system implementation and evaluation, proposing actionable recommendations and future directions to improve task-oriented conversational AI systems.

\subsection{Lessons Learned}
\label{sec:lessons}

\subsubsection{Leveraging Open-Source and Free Software}

An advantage of developing conversational systems today is the availability of reliable open-source and free software. The modern capabilities of these tools, especially in terms of speed and quality, make it possible to create complex, high-performance systems without costly licensing fees. Every component in our system -- from environment assets and avatars, to generative text and audio models -- was built using tools that are either open-source or free to use. While paid APIs often produce higher-quality output, they come with their own limitations of potential downtime, higher latency, and recurring costs. For projects where hardware capabilities allow, we recommend exploring locally-deployed alternatives. These not only reduce dependency on external services but also enable greater control over the system's responsiveness and reliability. As consumer hardware improves and demand for conversational applications grows, we anticipate further advancements in open-source tools, creating a rich ecosystem with plenty of fast and quality options to choose from.

\subsubsection{Avatars}  

Avatars are central to creating an immersive experience, and our current implementation revealed areas for improvement. Participants noted that the avatars appeared too cartoony, which diminished realism. We recommend using higher-fidelity models such as from the Rocketbox~\cite{rocketbox} or VALID~\cite{doValid2023} avatar libraries, and ensuring avatars turn toward the user based on proximity, as this feature was well-received. Future work will incorporate idle animations, such as subtle movements, to enhance realism and engagement further.

\subsubsection{Scenario Design}

Designing effective scenarios is crucial for user studies involving conversational AI, especially with multiple factors. We recommend using distinct but comparable scenarios to minimize bias from confounding variables and applying careful counterbalancing to account for order effects. Repeating the same scenario under different conditions, as in our pilot, introduced unintended behaviors like memorization or experimentation, which reduced engagement and added noise to the data~\cite{yu2020engaging}. Despite these issues, the quest-like, task-oriented approach proved effective overall, guiding participants naturally through interactions with virtual characters.

\subsection{Future Work}

\subsubsection{Gesture Recognition Integration}

Humans intuitively interpret nonverbal language such as gestures, and effectively use it to communicate in virtual social and collaborative settings~\cite{gahamndi2024collaboration, taranta2020ecovalid}. While our current implementation does not give virtual avatars the ability to see users' gestures, in future work we plan to employ a continuous (real-time) gesture recognizer such as Machete~\cite{taranta2021machete, taranta2022vkm} or OO-dMVMT~\cite{Cunico_2023_CVPR}, and appending a recognized gesture class to the message history of the nearest agent. An alternative recognition approach could involve passing screenshots from the virtual avatar's point-of-view to a visual language model (VLM), prompting it to classify gestures of an embodied human. Gestures could also trigger microphone input instead of pressing a dedicated controller button or pointing at the avatar~\cite{maslych2024selectionsdatabase}, reducing reliance on manual inputs. By making avatars more perceptive to nonverbal cues, this approach could improve the naturalness of interactions and create a more dynamic user experience~\cite{aneja_understanding_2021}.

\subsubsection{Response Delay Mitigation}

Generating speech responses is computationally intensive, requiring sequential processing through ASR, LLM and TTS systems. In our current architecture, the TTS engine relies on receiving the complete response text before generating audio, resulting in an average SRT of 3.2 seconds. In future work, we will explore token streaming to enable overlapping processing, allowing audio for subsequent sentences to be generated while earlier ones are still played.
Given the inherent latency, it would be useful to derive design recommendations to improve system usability. A promising direction is to mitigate perceived delays through conversational fillers, such as gesture or voice utterances, while responses being are generated. Prior work has shown that fillers can reduce perceived latency, but experiments were limited to pre-scripted interactions with fixed delays~\cite{kum_can_2022, gambino2017beyond} or Wizard-of-Oz setups~\cite{jeong2019exploring, funakoshi_smoothing_2008}. A useful experiment would apply capabilities and speed of modern models, investigating the effect of conversational fillers on perceived latency and learning outcomes in free-form conversations with IVAs.

\section{Conclusion}

In this work, we demonstrated the use of LLMs for conversational avatars in VR, exploring design considerations for response feedback and realism. By detailing two system architectures and incorporating user feedback, we provide practical insights to guide future development of task-oriented conversational AI systems. Additionally, we outlined promising directions for future work, including possible approaches to gesture recognition integration and response delay mitigation through token streaming and conversational fillers. These advancements aim to enhance the naturalness and efficiency of interactions, paving the way for more immersive and responsive virtual environments.


\bibliographystyle{ACM-Reference-Format}
\bibliography{main}

\appendix

\end{document}